\documentclass[reprint,amsmath,amssymb,aps,pra]{revtex4-2}
\usepackage[colorlinks,linkcolor=blue,anchorcolor=blue,citecolor=blue]{hyperref}
\usepackage{graphicx}
\usepackage{xcolor}
\usepackage{amsmath}
\usepackage{braket}

\begin{document}

\title{Nonlinear Landau-Zener Tunneling with Heat-Bath Colored Noise}
\author{Yi Cao}
\email[]{caoyi24@gscaep.ac.cn}
\author{Jie Liu}
\email[]{jliu@gscaep.ac.cn}
\affiliation{Graduate School, China Academy of Engineering Physics, Beijing 100193, China}
\date{\today}

\begin{abstract}
In this work, we investigate the influence of colored noise originating from a heat bath on nonlinear Landau-Zener (LZ) tunneling. The nonlinearity might arise intrinsically from many-body mean-field atomic self-interactions or Kerr nonlinearities in optical systems. Contrary to the well-established picture of thermally enhanced quantum tunneling within the linear LZ paradigm, ensemble statistics of our numerical simulations demonstrate that nonlinearity can give rise to thermally suppressed LZ tunneling: the tunneling probability of the nonlinear LZ system decreases as the amplitude of the colored noise increases beyond a critical value. Analytically, the Wiener-Hermite expansion facilitates elucidation of the intrinsic mechanism accounting for the nonmonotonic dependence of tunneling probability on noise amplitude. We compare these behaviors with stochastic resonance responses and clarify that the two phenomena stem from distinct physical mechanisms. In addition, we obtain approximate solutions in the weak- and strong-noise limits and discuss the implications of our theory.
\end{abstract}

\maketitle
\section{Introduction}
The Landau-Zener (LZ) model describes quantum transitions in systems swept through an avoided level crossing at a constant sweep rate~\cite{ShevchenkoPR2010, IvakhnenkoPR2023}, with widespread applications across diverse physical platforms, including superconducting qubits~\cite{IzmalkovEPL2004}, quantum dots~\cite{CaoNC2013}, cold atoms~\cite{SalgerPRL2007} and optical systems~\cite{LonghiLPR2009}, to name only a few. In realistic environments, quantum systems are inevitably coupled to heat baths, and the resulting relaxation and dephasing significantly modify the tunneling dynamics~\cite{Weiss2021, Breuer2007, GrifoniPR1998}. As such, LZ-based quantum-control protocols need to account for these heat-bath effects. The thermal effects have been extensively studied in LZ tunneling~\cite{Kayanuma1985, AoPRL1989, NalbachPRL2009, KrzywdaPRB2020, KrzywdaPRB2021, DaiNC2025}. Specifically, heat-bath-induced noise introduces stochastic fluctuations in the energy bias, which can drive incoherent thermal excitations across the energy gap~\cite{Kayanuma1984a, ChristieJPA2024}. This incoherent excitation mechanism acts alongside coherent tunneling and can enhance the tunneling probability.

The nonlinear LZ model~\cite{WuPRA2000, ZobayPRA2000, LiuPRA2002, WuPRL2006, TrimbornNJP2010, CaoPRA2023, CaoPRA2024} is a timely and valuable extension of the standard LZ framework, in which the energy difference depends on the populations. The natural occurrence of this nonlinearity through many-body mean-field self-interactions in ultracold atomic systems~\cite{LeggettRMP2001} and Kerr nonlinearities in optical systems~\cite{LedererPR2008, KhomerikiPRA2010} has made the model a subject of considerable interest~\cite{IvakhnenkoPR2023}. However, the effects of thermal noise on such nonlinear systems remain elusive.

In this work, we investigate the effects of colored noise from a heat bath on nonlinear LZ tunneling. We compute the ensemble-averaged tunneling probability as a function of noise amplitude and correlation time for different nonlinear parameters. For linear systems, the tunneling probability increases monotonically with noise amplitude. By contrast, nonlinearity fundamentally alters this behavior: the tunneling probability rises initially, peaks at a finite noise amplitude, and then decreases before converging to a strong-noise limit that is independent of nonlinear parameters. We further demonstrate that the locations of the tunneling probability peaks depend on the correlation time of colored noise. Analytically, the Wiener-Hermite expansion clarifies the intrinsic physical mechanism underlying the nonmonotonic variation with noise amplitude and delivers approximate analytical results in both the weak- and strong-noise regimes. We also compare the observed nonmonotonic tunneling behavior with stochastic resonance and confirm that the two phenomena arise from distinctly different physical mechanisms.

The remainder of this paper is organized as follows. Section~\ref{Sec-Model} introduces the stochastic nonlinear LZ model and presents the numerical results. In Sec.~\ref{Sec-WHE}, we derive the Wiener-Hermite hierarchy and its fast-noise reduction. Section~\ref{Sec-Phase-Diagram} discusses the weak- and strong-noise approximations and obtains a phase diagram. Section~\ref{Sec-Conclusion} summarizes the results and discusses the experimental implications. Appendices~\ref{Sec-Appendix1} and \ref{Sec-Appendix2} provide the derivations of the effective stochastic Hamiltonian and the tunneling probability in the strong-noise limit, respectively.

\section{Model and Numerical Simulations}\label{Sec-Model}
\subsection{Stochastic Schr\"odinger Equation}
In this work, we investigate a nonlinear LZ model coupled to a bosonic heat bath, where the system nonlinearity arises intrinsically from the many-body mean-field self-interaction~\cite{LeggettRMP2001, SmerziPRL1997, ZobayPRA2000}. The total Hamiltonian is decomposed as $H=H_{S}+H_{B}+H_{I}$, with the system part $H_{S}$ given by $H_{S}=\frac{1}{2}v\sigma_{x}+\frac{1}{2}\left[\alpha t+c(|b|^{2}-|a|^{2})\right]\sigma_{z}$~\cite{WuPRA2000, LiuPRA2002, WuPRL2006}. Here, $\alpha$ is the sweep rate of the energy bias across the avoided crossing, $v$ is the coupling strength between the two levels, and $c$ characterizes the nonlinear self-interaction that introduces population-dependent energy shifts. $(a, b)^{T}$ represents the two-mode wave function. $\sigma_{x}$ and $\sigma_{z}$ are Pauli matrices. Since the Hamiltonian can be scaled by dividing by $v$, for convenience, we can set $v=1$ as the energy unit hereafter. The heat bath is modeled as a collection of independent quantum harmonic oscillators, with Hamiltonian $H_{B}=\sum_{k}\left(\frac{p_{k}^{2}}{2}+\frac{1}{2}\omega_{k}^{2}q_{k}^{2}\right)$~\cite{FordJMP1965, Gardiner2004}, where $q_{k}$ and $p_{k}$ are position and momentum operators of the $k$-th bath oscillator, and $\omega_{k}$ is the corresponding angular frequency. Following the Caldeira-Leggett framework, we adopt the system-bath coupling~\cite{Kayanuma1984a, LeggettRMP1987}: $H_{I}=\frac{1}{2}\sum_{k}\eta_{k}q_{k}\sigma_{z}$, where $\eta_{k}$ is the coupling constant between the bath and the system.

To proceed with the derivation, we assume a Drude spectral density for the heat bath, $J(\omega)=\eta\omega\gamma/(\omega^{2}+\gamma^{2})$~\cite{Breuer2007, Weiss2021}, where $\eta$ denotes the system-bath coupling and $\gamma$ is the Drude cutoff frequency. In the classical-noise limit, the fluctuating force exerted by the heat bath is represented by a real Gaussian stochastic process~\cite{FordPRA1988}. Within the Markov approximation, the nonlocal memory term is reduced to a time-local renormalization, which is incorporated into the system Hamiltonian~\cite{Gardiner2004}. Solving the bath dynamics and applying these approximations then yield the effective stochastic Hamiltonian
\begin{equation}\label{Eq-Effective_Hamiltonian}
    H_{\text{eff}}=\frac{1}{2}\begin{pmatrix} \mathcal{L}(t) & v \\ v & -\mathcal{L}(t)
    \end{pmatrix},
\end{equation}
where $\mathcal{L}(t)=\alpha t+c(|b|^2-|a|^2)+\xi(t)$. The derivation is detailed in Appendix~\ref{Sec-Appendix1}. The dynamics of the two-mode wave function are governed by the Schr\"odinger equation $i\frac{d}{d t}\begin{pmatrix} a \\ b \end{pmatrix}=H_{\text{eff}}\begin{pmatrix} a \\ b \end{pmatrix}$, with $\hbar=1$, which preserves the normalization $|a|^{2}+|b|^{2}=1$. Here, $\xi(t)$ denotes a zero-mean colored Gaussian noise with an exponentially decaying temporal correlation:
\begin{equation}\label{Eq-Colored_Noise}
    \langle \xi(t) \rangle=0,\quad \langle \xi(t)\xi(s) \rangle=D^{2}e^{-\gamma|t-s|},
\end{equation}
where $D=\sqrt{\eta k_{B}T}$ is the noise amplitude, with $k_{B}$ being the Boltzmann constant and $T$ the temperature of the heat bath, and $1/\gamma$ is the noise correlation time. This colored noise can be realized as an Ornstein-Uhlenbeck process, governed by the stochastic differential equation~\cite{Gardiner1986}
\begin{equation}\label{Eq-SDE}
    d\xi(t)=-\gamma \xi(t)+\sqrt{2\gamma}D dW(t),
\end{equation}
where $dW(t)$ is a real Wiener increment satisfying $\langle dW(t) \rangle=0$ and $\langle [dW(t)]^{2} \rangle=dt$, with independent increments. 

\subsection{Ensemble Statistics}
\begin{figure}[t]
\centering
\includegraphics[width=\linewidth]{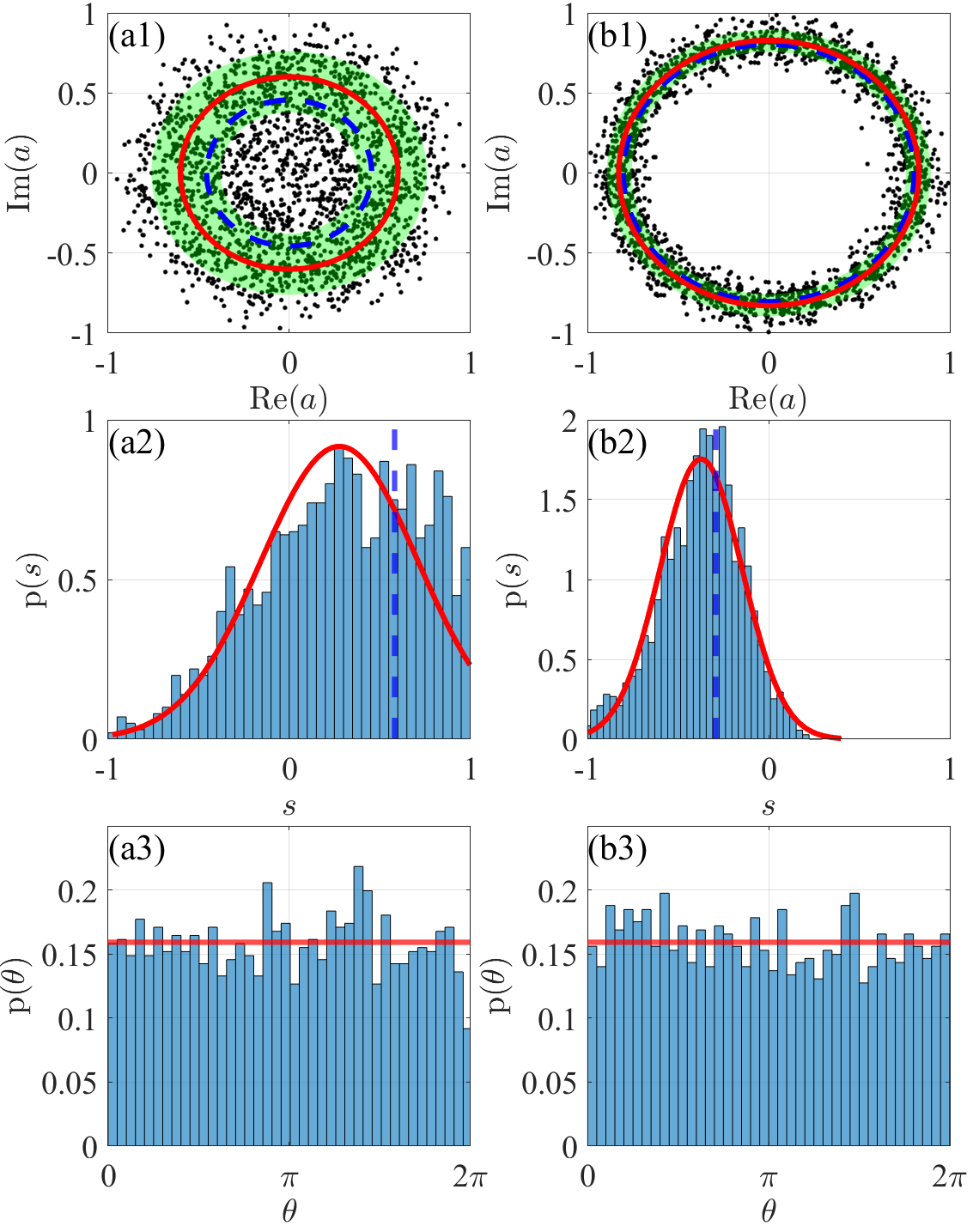}
\caption{Long-time ensemble statistics of the probability amplitude $a$, population difference $s$, and relative phase $\theta$ in the linear [panels (a1)--(a3), $c=0$] and nonlinear [panels (b1)--(b3), $c=5$] systems. (a1, b1) Black dots show the long-time values of $a$ for the individual noise realizations in the complex-$a$ plane; the red solid line and shaded band indicate the ensemble mean and its statistical fluctuations, respectively. (a2, b2) Probability densities of $s$, with red solid curves denoting Gaussian fits. (a3, b3) Probability densities of $\theta$, with red solid curves denoting the uniform distribution with amplitude $1/(2\pi)$. Blue solid lines in (a1, b1) and (a2, b2) show the noiseless results. Parameters: $\alpha=1$, $D=1$, and $\gamma=1$.}
\label{Fig-Figure1}
\end{figure}
We now perform numerical calculations of the tunneling probability. Specifically, we numerically solve the stochastic Schr\"odinger equation using the fourth-order Runge-Kutta method. The noise variable $\xi(t)$ is obtained by solving Eq.~(\ref{Eq-SDE}) via the Heun scheme~\cite{Gard1988}. To initialize the noise variable for ensemble simulations, we consider the time-dependent probability distribution $\Phi(\xi,t)$ of the noise variable. Its time evolution follows the deterministic Fokker-Planck equation~\cite{Gardiner1986}, $\partial_{t}\Phi(\xi,t)=\gamma\partial_{\xi}[\xi\Phi(\xi,t)]+\gamma D^{2}\partial_{\xi}^{2}\Phi(\xi,t)$. Imposing the steady-state condition $\partial_{t}\Phi(\xi,t)=0$ gives rise to the stationary Gaussian distribution
\begin{equation}\label{Eq-Steady-state distribution}
    \Phi_{\text{st}}(\xi)=\frac{1}{\sqrt{2\pi D^{2}}}\exp\left(-\frac{\xi^{2}}{2D^{2}}\right).
\end{equation}
For each noise realization, the initial value of $\xi$ is sampled from $\Phi_{\text{st}}(\xi)$ to ensure thermal-equilibrium initialization.

For sufficiently large $|t|$, the sweep term $\alpha t$ dominates the noisy, nonlinear, and off-diagonal contributions to the Hamiltonian in Eq.~(\ref{Eq-Effective_Hamiltonian}). The instantaneous eigenstates therefore approach the diabatic states $(1,0)^{T}$ and $(0,1)^{T}$, with eigenvalues $\alpha t/2$ and $-\alpha t/2$, respectively. Each realization is initialized in $(a,b)^{T}=(1,0)^{T}$. At long times, the state is characterized by the population difference $s=|b|^{2}-|a|^{2}$ and relative phase $\theta=\theta_{b}-\theta_{a}$, where $a=|a|e^{i\theta_{a}}$ and $b=|b|e^{i\theta_{b}}$. The corresponding tunneling probability is $P=(1-s)/2$. Figure~\ref{Fig-Figure1} summarizes the long-time results from $M=2000$ independent noise realizations. Without noise, the probability amplitude $a$ lies on a circle of fixed radius in the complex plane, whereas noise spreads the final amplitudes over an extended region and increases the ensemble-averaged magnitude of $a$. Under the same noise conditions, the linear system ($c=0$) exhibits a noticeably broader spread of final amplitudes than the nonlinear system ($c=5$), as shown in panels (a1) and (b1). The corresponding population-difference histograms in panels (a2) and (b2) are approximately Gaussian, with a substantially larger width in the linear case. In contrast, the relative-phase distributions in panels (a3) and (b3) become nearly uniform on $[0,2\pi]$. These results show that colored noise induces pronounced variations in both population and phase across individual noise realizations, while nonlinearity suppresses the population fluctuations.

\subsection{Nonmonotonic Landau-Zener tunneling probability}
The long-time distributions in Fig.~\ref{Fig-Figure1} show that the tunneling probability varies from one noise realization to another. We therefore quantify the tunneling dynamics by the ensemble-averaged probability,
\begin{equation}
    \bar{P}=\frac{1}{M}\sum_{i=1}^{M}P_{i},
\end{equation}
where $P_i$ is the long-time tunneling probability obtained from the $i$th realization and $M$ is the number of noise realizations. All numerical results presented below are obtained by averaging over $M=2000$ independent noise realizations. Numerical convergence was verified by doubling the sample size.
\begin{figure}[t]
\centering
\includegraphics[width=\linewidth]{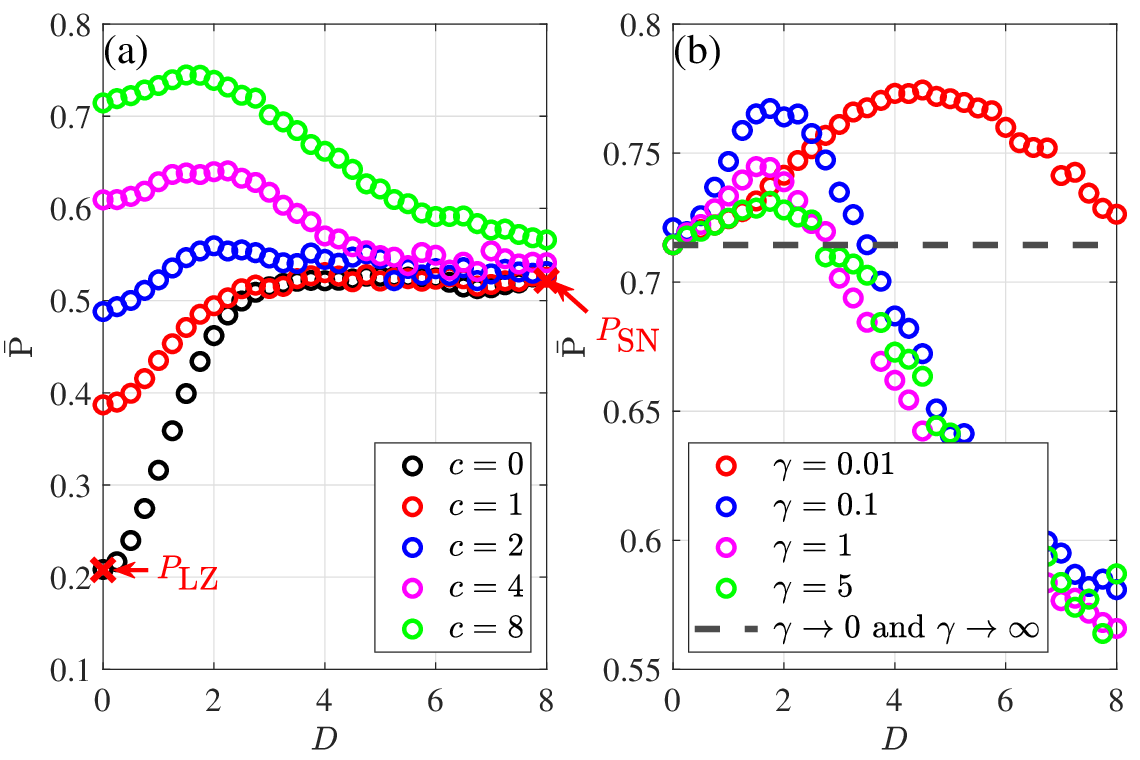}
\caption{Ensemble-averaged tunneling probability $\bar{P}$ as a function of $D$ for $\alpha=1$. (a) We choose a typical correlation time parameter $\gamma=5$ and show nonmonotonic LZ tunneling probability vs. $D$ when $c\neq0$; (b) We set the nonlinear parameter $c=8$ and show the correlation time dependence of the tunneling probability peaks. When $\gamma\to0$ and $\gamma\to\infty$, the tunneling probabilities turn to be independent of the noise and the tunneling peaks completely disappear.}
\label{Fig-Figure2}
\end{figure}

For the linear system ($c=0$), the ensemble-averaged tunneling probability $\bar{P}$ in Fig.~\ref{Fig-Figure2}(a) increases monotonically with the noise amplitude $D$, from the noiseless LZ value $P_{\mathrm{LZ}}=\exp\left[-\pi v^2/(2\alpha)\right]$ toward the strong-noise limit $P_{\mathrm{SN}}=\frac{1}{2}\left[1+\exp(-\pi v^2/\alpha)\right]$. This behavior has been investigated in Ref.~\cite{Kayanuma1984a}. Physically, near the avoided crossing, thermal fluctuations induce transitions from the lower to the upper eigenstate, thereby increasing $\bar{P}$. In the nonlinear counterpart, the tunneling probability at $D=0$ exceeds that of the linear system because the nonlinear self-interaction accelerates the sweep process through an effective rate $\tilde{\alpha}=\alpha+2c(v/2)^2\sqrt{\pi/\tilde{\alpha}}>\alpha$, as reported in Ref.~\cite{LiuPRA2002}. In the strong-noise limit $D\to\infty$, we find that the tunneling probability converges to $P_{\mathrm{SN}}$, independent of the nonlinear parameter $c$. At intermediate noise amplitudes, the tunneling probability $\bar{P}$ varies nonmonotonically with noise amplitude $D$: it initially increases, reaches a maximum at finite $D$, and then decreases to $P_{\mathrm{SN}}$. In the simulations above, we set the sweep rate to unity. Nonetheless, our findings hold for any finite sweep rate. This is because the characteristic LZ timescale is $1/\sqrt{\alpha}$, which provides a natural temporal scale for examining the sweep-rate dependence. The noise amplitude and correlation rate can then be compared with the corresponding scale $\sqrt{\alpha}$ through the ratios $D/\sqrt{\alpha}$ and $\gamma/\sqrt{\alpha}$~\cite{CaoArXiv2026, Kayanuma1984b}, respectively.

Notably, the peak structures we observed in Fig.~\ref{Fig-Figure2}(a) are very similar to those of stochastic resonance~\cite{McNamaraPRA1989}. In conventional stochastic resonance, a weak periodic drive modulates a bistable potential, while white noise induces interwell transitions at the Kramers rate, and the response is maximal when the mean residence time matches half the driving period~\cite{GammaitoniRMP1998}. However, in our system, there is no periodic drive and we find that the locations of the tunneling probability peaks depend on the correlation time of the noise, as shown in Fig.~\ref{Fig-Figure2}(b). In the slow-noise limit $\gamma\to0$, the noise merely shifts the time at which the crossing occurs and therefore does not alter the transition probability. In the white-noise limit $\gamma\to\infty$, according to Eq.~(\ref{Eq-Colored_Noise}), $\langle \xi(t)\xi(s) \rangle\simeq\frac{D^{2}}{\gamma}\delta(t-s)$, which implies that the effective noise strength $D^2/\gamma$ vanishes in this case. Consequently, the peaks disappear in both limits, as shown in Fig.~\ref{Fig-Figure2}(b). 

In the following sections, we use the Wiener-Hermite expansion to help us to understand the underlying mechanism of the distinct tunneling peaks.

\section{Wiener-Hermite Expansion}\label{Sec-WHE}
In this section, we begin with the stochastic Liouville equation $id\rho/dt=[H_{\text{eff}},\rho]$. For each realization of the Ornstein-Uhlenbeck process, we introduce the Bloch variables $X_{\xi}=\rho_{12}+\rho_{21},~Y_{\xi}=-i(\rho_{12}-\rho_{21})$, and $Z_{\xi}=\rho_{11}-\rho_{22}$. Their dynamics is governed by
\begin{subequations}
    \begin{align}
        \dot X_\xi&=(\alpha t-cZ_\xi+\xi)Y_\xi,\\
        \dot Y_\xi&=-(\alpha t-cZ_\xi+\xi)X_\xi+vZ_\xi,\\
        \dot Z_\xi&=-vY_\xi,
    \end{align}
\end{subequations}
where the subscript $\xi$ emphasizes the dependence on the complete noise history. To obtain a deterministic description, we introduce the conditional averages $\bar{U}(t,\xi)=\mathbb{E}[U_\xi(t)\mid\xi(t)=\xi]$ for $U=X,Y,Z$. With the Ornstein-Uhlenbeck generator $\mathcal L_{\mathrm{OU}}=-\gamma\xi\partial_{\xi}+\gamma D^2\partial_{\xi}^{2}$, the conditional equations read $\partial_t\bar X=(\alpha t+\xi)\bar{Y}-c(\bar{Z}\bar{Y}+C_{ZY})+\mathcal L_{\mathrm{OU}}\bar{X},~\partial_t\bar{Y}=-(\alpha t+\xi)\bar X+c(\bar Z\bar X+C_{ZX})+v\bar{Z}+\mathcal L_{\mathrm{OU}}\bar{Y},$ and $\partial_t\bar{Z}=-v\bar{Y}+\mathcal{L}_{\mathrm{OU}}\bar{Z}$. Here, $C_{ZY}=\mathbb{E}[Z_{\xi}Y_{\xi}\mid\xi]-\bar{Z}\bar{Y}$ and $C_{ZX}=\mathbb{E}[Z_{\xi}X_{\xi}\mid\xi]-\bar{Z}\bar{X}$ are conditional covariances. Their equations couple to conditional moments of successively higher order and therefore generate an infinite hierarchy~\cite{Gardiner1986}. Within a moment-closure approximation~\cite{SchnoerrJPA2017}, we set $C_{ZY}=C_{ZX}=0$ and denote the resulting conditional means by $X(t,\xi)$, $Y(t,\xi)$, and $Z(t,\xi)$. They obey
\begin{subequations}
    \begin{align}
        \partial_tX&=(\alpha t-cZ+\xi)Y+\mathcal L_{\mathrm{OU}}X,\\
        \partial_tY&=-(\alpha t-cZ+\xi)X+vZ+\mathcal L_{\mathrm{OU}}Y,\\
        \partial_tZ&=-vY+\mathcal L_{\mathrm{OU}}Z.
    \end{align}
\end{subequations}
At each fixed time, the conditional means $X(t,\xi)$, $Y(t,\xi)$, and $Z(t,\xi)$ are expanded in the normalized Hermite basis associated with the stationary Gaussian variable $\xi$~\cite{XiuSISC2002, Maitre2010, Cho2011}: $U(t,\xi)=\sum_{n=0}^{\infty}U_n(t)\Psi_n(\xi)$ for $U=X,Y,Z$, with $U_n(t)$ denoting the coefficient of the $n$-th Hermite mode, where $\Psi_n(\xi)=H_n(\xi/D)/\sqrt{n!}$ and $H_n(\cdot)$ denotes the $n$th probabilists' Hermite polynomial. The ensemble average is defined by $\mathbb E[F(\xi)]=\int_{-\infty}^{\infty}F(\xi)\Phi_{\mathrm{st}}(\xi)d\xi$, where $\Phi_{\mathrm{st}}(\xi)$ is the stationary probability density in Eq.~(\ref{Eq-Steady-state distribution}). Hence, $\mathbb{E}[\Psi_n\Psi_m]=\delta_{nm}$, $\mathcal{L}_{\mathrm{OU}}\Psi_n=-n\gamma\Psi_n$ and $\xi\Psi_{n}=D(\sqrt{n+1}\Psi_{n+1}+\sqrt{n}\Psi_{n-1})$~\cite{Xiu2010}. The nonlinear terms are projected through the Hermite triple-product tensor $M_{nij}=\mathbb{E}[\Psi_n\Psi_i\Psi_j]$. Its nonzero elements admit the closed-form expression $M_{nij}=\sqrt{i!j!n!}/[k!(i-k)!(j-k)!]$ with $k=(i+j-n)/2$, provided that $i+j+n$ is even and the triangle condition $|i-j|\le n\le i+j$ is satisfied. Otherwise, $M_{nij}=0$~\cite{Eigel2020}.

Applying the stochastic Galerkin projection~\cite{XiuSISC2002, Maitre2010}, we multiply each equation by $\Psi_n$ and average over the Gaussian measure, obtaining
\begin{widetext}
\begin{subequations}\label{Eq-WHE}
\begin{align}
    \frac{\partial X_{n}}{\partial t}=&-n\gamma X_{n}+\alpha t Y_{n}-c\sum_{i,j=0}^{\infty}M_{nij}Z_{i}Y_{j}+D\left(\sqrt{n+1}Y_{n+1}+\sqrt{n}Y_{n-1}\right),\\
    \frac{\partial Y_{n}}{\partial t}=&-n\gamma Y_{n}-\alpha t X_{n}+c\sum_{i,j=0}^{\infty}M_{nij}Z_{i}X_{j}+vZ_{n}-D\left(\sqrt{n+1}X_{n+1}+\sqrt{n}X_{n-1}\right),\\
    \frac{\partial Z_{n}}{\partial t}=&-n\gamma Z_{n}-vY_{n}.
\end{align}
\end{subequations}
\end{widetext}
When $c=0$, the above equations reduce to those in Ref.~\cite{Kayanuma1984b}. 

We focus on the fast-noise regime $\gamma\gg1$, where the noise correlation time $1/\gamma$ is much shorter than the timescale of the LZ transition. Under this condition, the higher-order Hermite modes with $n\ge1$ relax rapidly. Within the steady-state approximation, we neglect modes with $n\ge2$, retain the leading $n=1$ modes, and set their time derivatives to zero, obtaining $X_1\simeq(D/\gamma)Y_0$ and $Y_1\simeq-(D/\gamma)X_0$. Substitution into the $n=0$ equations yields
\begin{subequations}
\begin{align}  
    \dot{X}_{0}=&(\alpha t-cZ_{0})Y_{0}-\Gamma X_{0},\\
    \dot{Y}_{0}=&-(\alpha t-cZ_{0})X_{0}+vZ_{0}-\Gamma Y_{0},\\
    \dot{Z}_{0}=&-vY_{0},
\end{align}
\end{subequations}
where $\Gamma=D^2/\gamma$ is the effective decoherence rate. These equations can also be derived from the Lindblad master equation~\cite{TrimbornNJP2010}. We now specify the initial conditions as follows. The system is prepared in state $(a,b)^{T}=(1,0)^{T}$, corresponding to $X_{n}(t\to-\infty)=Y_{n}(t\to-\infty)=0$ for all $n\ge0$, $Z_{0}(t\to-\infty)=1$, and $Z_{n}(t\to-\infty)=0$ for all $n\ge1$. The ensemble-averaged tunneling probability is given by $\bar{P}=\lim_{t\to\infty}\frac{1+Z_{0}(t)}{2}$.

\section{Approximate Solutions and Phase Diagram}\label{Sec-Phase-Diagram}
For the strong-noise limit, we start from the effective fast-noise equations derived above. In the regime $\Gamma\gg\sqrt{\alpha}$, the transverse components $X_{0}$ and $Y_{0}$ relax on the timescale $1/\Gamma$, much faster than the population difference $Z_{0}$. Adiabatically eliminating $X_{0}$ and $Y_{0}$ and introducing the rescaled time $\tau=t/\Gamma$ together with $\epsilon=c/\Gamma$, we obtain
\begin{equation}
    \frac{d Z_{0}}{d \tau}=-\frac{v^2}{1+(\alpha \tau-\epsilon Z_{0})^{2}}Z_{0}.
\end{equation}
When $\Gamma\gg c$, the parameter $\epsilon$ is small, and the nonlinear term can be treated perturbatively. A first-order expansion in $\epsilon$ (see Appendix~\ref{Sec-Appendix2}) yields
\begin{equation}\label{Eq-Phase-Strong}
    \begin{split}
    \bar{P}\simeq&\frac{1}{2}+\frac{1}{2}e^{-\frac{\pi v^{2}}{\alpha}}+\frac{c}{\Gamma}\frac{v^{2}/\alpha}{(v^{2}/\alpha)^{2}+4}\\
    &\times\left(e^{-\frac{\pi v^{2}}{\alpha}}-e^{-\frac{2\pi v^{2}}{\alpha}}\right).
    \end{split}
\end{equation}
In the above equation, the first term, $1/2$, is the equal-population result of complete mixing. The second term, $\frac{1}{2}\exp(-\pi v^{2}/\alpha)=P_{\mathrm{LZ}}^{2}/2$, represents the residual population imbalance after the finite sweep~\cite{CaoArXiv2026}. The third term represents the nonlinear correction. Consequently, all curves in Fig.~\ref{Fig-Figure2}(a) approach the same $c$-independent strong-noise value $P_{\mathrm{SN}}=[1+\exp(-\pi v^{2}/\alpha)]/2$ as $\Gamma\to\infty$.

In this limit, the strong noise rapidly destroys the coherence between the two levels, and population transfer is governed by noise-induced excitations. Moreover, we find that the nonlinearity is effectively reduced by the colored noise with a scaling of $c/\Gamma$, which implies the nonlinear acceleration mechanism mentioned in Sec.~\ref{Sec-Model}C is weakened by the noise. In Bose-Einstein condensate (BEC) systems of ultracold atoms, this means that the superfluidity can be destroyed by strong noise resulting in the reduction of the tunneling probability~\cite{SnizhkoPRA2016, KumarPRA2017}.

For the linear LZ model with $c=0$, we observe that the noise-modulated LZ tunneling in Fig.~\ref{Fig-Figure2}(a) can be approximately expressed by $P_{\mathrm{SN}}-\left(P_{\mathrm{SN}}-P_{\mathrm{LZ}}\right)e^{-2\Gamma}$. In the weak-noise limit of $\Gamma\ll1$, the nonlinear effects will accelerate the sweep process through an effective rate $\tilde{\alpha}=\alpha+2c(v/2)^2\sqrt{\pi/\tilde{\alpha}}$~\cite{LiuPRA2002}, i.e., $\bar{P}\simeq P_{\mathrm{SN}}(\tilde{\alpha})-\left(P_{\mathrm{SN}}(\tilde{\alpha})-P_{\mathrm{LZ}}(\tilde{\alpha})\right)e^{-2\Gamma}$. Expanding $\bar{P}$ to first order in $\Gamma$, we have
\begin{equation}\label{Eq-Phase-Weak}
    \bar{P}\simeq\exp\left(-\frac{\pi v^{2}}{2\tilde{\alpha}}\right)+\Gamma\left[1-\exp\left(-\frac{\pi v^{2}}{2\tilde{\alpha}}\right)\right]^{2}.
\end{equation}
The first term is just the nonlinear LZ tunneling probability $P_{\mathrm{NLZ}}$. The second term, which is linear in $\Gamma$, indicates the noise-induced incoherent thermal excitations.

From Eq.~(\ref{Eq-Phase-Weak}), $\bar{P}$ increases with $\Gamma$, indicating that the coherent quantum tunneling dominates in the weak-noise regime, and the thermal excitations assist the tunneling probability because the noise can generate a number of additional avoided level crossings. 

However, in the strong-noise regime, Eq.~(\ref{Eq-Phase-Strong}) shows that $\bar{P}$ turns to decrease with increasing $\Gamma$. Therefore, as the noise amplitude increases, the tunneling probability is expected to exhibit a nonmonotonic profile. The maximum tunneling probability marks a transition from a coherent quantum-tunneling-dominated regime to a thermal-excitation-dominated regime. The critical noise amplitude $D_{\mathrm{cri}}$, which coincides with the peak position, may be estimated by comparing the expressions of $\bar{P}$ in the weak- and strong-noise limits.
\begin{figure}[t]
\centering
\includegraphics[width=\linewidth]{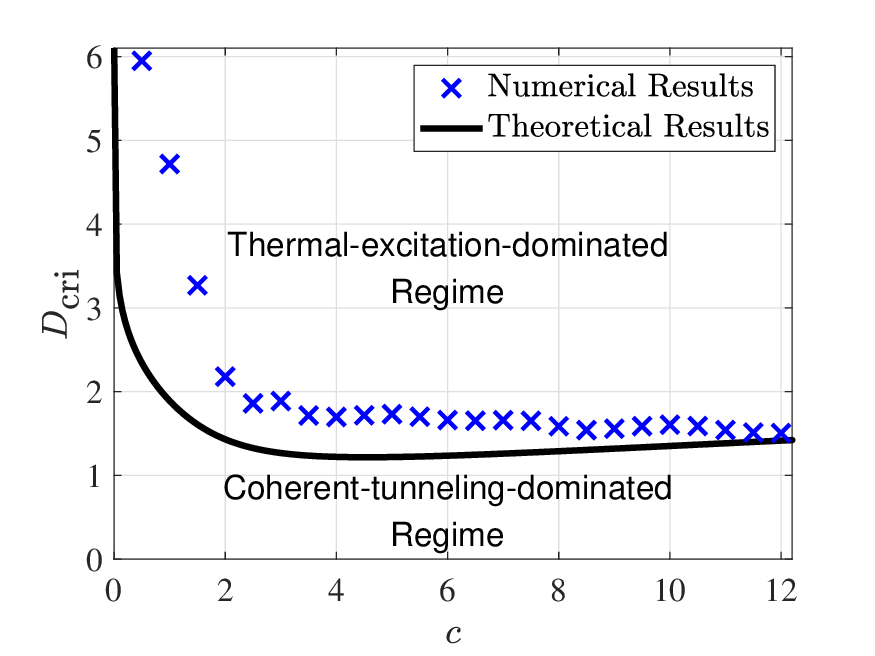}
\caption{The critical noise amplitudes $D_{\mathrm{cri}}$ delineate the boundary between the coherent-quantum-tunneling-dominated and thermal-excitation-dominated regimes of nonlinear LZ tunneling. The black solid line represents the theoretical prediction, and the blue markers denote the numerical results. Here, we choose the typical parameters $\alpha=1$ and $\gamma=5$.}
\label{Fig-Figure3}
\end{figure}

Figure~\ref{Fig-Figure3} presents the phase diagram describing the transition from the coherent quantum tunneling dominated regime to the thermally excited dominated regime. The theoretical predictions capture the overall trend observed in numerical simulations, despite existing quantitative discrepancies. The expression of $\bar{P}$ derived in the strong-noise limit is built upon the fast-noise condition $\gamma\gg1$, whereas the numerical calculations shown in Fig.~\ref{Fig-Figure3} employ a finite $\gamma$. Furthermore, the weak-noise-limit expression for $\bar{P}$ yields quantitatively reliable results only when $\Gamma$ is sufficiently small, yet the peak of the tunneling probability can fall beyond this valid parameter regime. These inherent limitations of the two approximations account for the deviations between theoretical and numerical results. In addition, the larger deviations between theoretical predictions and numerical observations emerge for small nonlinear parameters ($c<2$). This may partly arise from the fact that the expression for the effective sweep rate yields higher accuracy at relatively large nonlinearity, as addressed in Ref.~\cite{LiuPRA2002}.

\section{Conclusion}\label{Sec-Conclusion}
In conclusion, we have studied nonlinear LZ tunneling under heat-bath-induced colored noise using an effective stochastic Schr\"odinger equation. We show that the tunneling probability peaks at a finite noise amplitude, an intriguing behavior much analogous to stochastic resonance. This phenomenon is governed by nonlinearity and universal with respect to both system parameters and noise correlation time. The analytical Wiener-Hermite expansion uncovers the intrinsic mechanism accounting for the nonmonotonic dependence of tunneling probability on noise amplitude, and clarifies that our findings and stochastic resonance stem from distinct physical mechanisms. The phase diagram and the corresponding critical noise amplitudes for the transition from the coherent quantum-tunneling-dominated nonlinear LZ tunneling regime to thermal-excitation-dominated nonlinear LZ tunneling regime are obtained.

Experimentally, nonlinear LZ tunneling has been realized in BECs by accelerating an optical lattice so that the condensate is swept through an avoided crossing between Bloch bands, and atom-atom interactions provide the nonlinearity~\cite{MorschPRL2001, CristianiPRA2002, Jona-LasinioPRL2003, MorschRMP2006}. The present model can also be extended to Kerr-nonlinear waveguide arrays, where the paraxial wave equation assumes the form of a Schr\"odinger equation, with the propagation coordinate $z$ playing the role of time and waveguide curvature providing an optical analogue of LZ tunneling~\cite{LonghiLPR2009, DreisowPRA2009}. Colored noise can be introduced through a stochastic modulation of the applied transverse bias. Furthermore, a magnetic counterpart arises in molecular magnets, where field sweeps drive resonant spin tunneling and produce quantum steps in magnetization hysteresis loops. Accounting for dipolar interactions, these step heights exhibit a scaling with sweep rate and, together with the sample geometry, determine the single-molecule tunnel splitting~\cite{LiuPRB2002}. Controlled colored fluctuations of the longitudinal magnetic field would then allow the predicted noise-modified tunneling to be tested through these quantum steps. Our results deepen the physical understanding of the combined effects of nonlinearity and colored noise on LZ tunneling dynamics. The present work also provides useful guidance for quantum control and quantum simulation, and may stimulate further experimental studies in the platforms discussed above.

\begin{acknowledgments}
This work was supported by the Science Challenge Project (Grant No.~TZ2025017) and the National Natural Science Foundation of China (Grant No.~U2330401).
\end{acknowledgments}

\appendix
\section{Derivation of the Effective Hamiltonian in Eq.~(\ref{Eq-Effective_Hamiltonian})}\label{Sec-Appendix1}
In this appendix, we outline the derivation of the effective Hamiltonian (\ref{Eq-Effective_Hamiltonian}) presented in the main text. We set $S\equiv\frac{1}{2}\sigma_{z}$ for notational convenience. The Heisenberg equations of motion for the bath oscillators $\dot{q}_{k}=p_{k},~\dot{p}_{k}=-\omega_{k}^{2}q_{k}-\eta_{k}S$ are formally solved to give
\begin{equation}\label{Eq-General_Solution}
    q_{k}(t)=q_{k}^{(h)}(t)-\eta_{k}\int_{0}^{t}dt^{\prime}\frac{\sin\left[\omega_{k}(t-t^{\prime})\right]}{\omega_{k}}S(t^{\prime}),
\end{equation}
with the homogeneous part $q_{k}^{(h)}(t)=q_{k}(0)\cos\omega_{k}t+p_{k}(0)\frac{\sin(\omega_{k}t)}{\omega_{k}}$, which is determined solely by the initial coordinate $q_{k}(0)$ and momentum $p_{k}(0)$ of the bath oscillators. Substituting Eq.~(\ref{Eq-General_Solution}) into the Heisenberg equation for an arbitrary system operator $O$, we obtain
\begin{equation}\label{Eq-Heisenberg_Equation}
\begin{split}
    \dot{O}(t)=&\frac{1}{i\hbar}[O(t),H_{S}]+\frac{1}{i\hbar}[O(t),S]\xi(t)\\
    &-\frac{1}{i\hbar}[O(t),S]\int_{0}^{t}dt^{\prime}K(t-t^{\prime})S(t^{\prime}),
\end{split}
\end{equation}
where $\xi(t)=\sum_{k}\eta_{k}q_{k}^{(h)}(t)$ is the noise operator and $K(t)=\sum_{k}\frac{\eta_{k}^{2}}{\omega_{k}}\sin(\omega_{k}t)$ the memory kernel. We assume that the system and bath are initially decoupled, and the bath is initially in thermal equilibrium at temperature $T$, as described by the density matrix $\rho_{B}(0)=\frac{e^{-H_{B}/(k_{B}T)}}{\text{Tr}e^{-H_{B}/(k_{B}T)}}$, where $k_{B}$ is the Boltzmann constant. One can readily verify that $\langle \xi(t) \rangle=0$. This vanishing mean value follows directly from the thermal equilibrium distribution and the linear dependence of $\xi(t)$ on the bath coordinates and momenta. The correlation function of $\xi(t)$ is given by $\langle \xi(t)\xi(t^{\prime}) \rangle=\sum_{k}\frac{\eta_{k}^{2}\hbar}{2\omega_{k}}\big\{\coth\left(\frac{\hbar\omega_{k}}{2k_{B}T}\right)\cos[\omega_{k}(t-t^{\prime})]-i\sin\left[\omega_{k}(t-t^{\prime})\right]\big\}$~\cite{FordPRA1988}. In the continuum limit for the bath oscillators, we introduce the spectral density $J(\omega)=\frac{\pi}{2}\sum_{k}\frac{\eta_{k}^{2}}{\omega_{k}}\delta(\omega-\omega_{k})$~\cite{Breuer2007}. The correlation function can then be expressed as a continuous integral
\begin{equation}
\begin{split}
    \langle \xi(t)\xi(t^{\prime}) \rangle=\frac{\hbar}{\pi}\int_{0}^{\infty}d\omega J(\omega)\bigg\{&\coth\left(\frac{\hbar\omega}{2k_{B}T}\right)\cos\left[\omega(t-t^{\prime})\right]\\
    &-i\sin\left[\omega(t-t^{\prime})\right]\bigg\}.
\end{split}
\end{equation}
In what follows, we choose the Drude spectral density $J(\omega)=\eta\omega\gamma/(\omega^{2}+\gamma^{2})$~\cite{Breuer2007, Weiss2021}, where $\eta$ is the effective system-bath coupling strength and $\gamma$ denotes the Drude cutoff frequency. In the classical limit $\hbar\to0$, employing the approximation $\coth\left(\frac{\hbar\omega}{2k_{B}T}\right)\approx\frac{2k_{B}T}{\hbar\omega}$, the correlation function of $\xi(t)$ reduces to $\langle \xi(t)\xi(t^{\prime}) \rangle=\eta k_{B}Te^{-\gamma|t-t^{\prime}|}-i\frac{\hbar\eta\gamma}{2}e^{-\gamma|t-t^{\prime}|}$. As $\hbar\to0$, the imaginary term vanishes and only the real exponential correlation survives. Consequently, $\xi(t)$ can be treated as classical colored Gaussian noise. This classical noise satisfies
\begin{equation}
    \langle \xi(t) \rangle=0,\quad \langle \xi(t)\xi(t^{\prime}) \rangle=\eta k_{B}Te^{-\gamma|t-t^{\prime}|}.
\end{equation}
For the Drude spectrum, the memory kernel is $K(t)=\eta\gamma e^{-\gamma t}$ for $t>0$. In the Markovian approximation, the bath correlation time $1/\gamma$ is short compared with the characteristic time scale of the system dynamics, allowing $S(t')$ to be replaced by $S(t)$ inside the memory integral~\cite{Gardiner2004}. The resulting time-local renormalization is absorbed into the system Hamiltonian. The Heisenberg equation then becomes $\dot{O}=\frac{1}{i\hbar}[O(t),H_S]+\frac{1}{i\hbar}[O(t),S]\xi(t)$, which is generated by $H_{\mathrm{eff}}=H_S+\xi(t)S$ and is precisely Eq.~(\ref{Eq-Effective_Hamiltonian}) in the main text.

\section{The Derivation of Eq.~(\ref{Eq-Phase-Strong}) in the Strong-Noise Limit }\label{Sec-Appendix2}
This appendix derives the leading nonlinear correction in the strong-noise limit. For $\Gamma\gg\sqrt{\alpha}$, adiabatic elimination of the rapidly relaxing transverse components reduces the dynamics of $Z_{0}$ to $\frac{dZ_{0}}{dt}=-\frac{\Gamma v^2}{\Gamma^2+(\alpha t-cZ_{0})^2}Z_{0}$. To extract the nonlinear correction, we additionally require $c/\Gamma\ll1$. Upon introducing $\tau=t/\Gamma$ and $\epsilon=c/\Gamma$, the equation becomes
\begin{equation}
    \frac{dZ_{0}}{d\tau}=-\frac{v^2}{1+(\alpha\tau-\epsilon Z_{0})^2}Z_{0},
    \qquad Z_{0}(-\infty)=1.
\end{equation}
We then expand the population difference as $Z_{0}=Z^{(0)}+\epsilon Z^{(1)}+\mathcal{O}(\epsilon^2)$. Equating terms of the same order in $\epsilon$ gives
\begin{subequations}
\begin{align}
    \frac{dZ^{(0)}}{d\tau}&=-\frac{v^2}{1+\alpha^2\tau^2}Z^{(0)},\\
    \frac{dZ^{(1)}}{d\tau}+\frac{v^2}{1+\alpha^2\tau^2}Z^{(1)}
    &=-\frac{2v^2\alpha\tau}{(1+\alpha^2\tau^2)^2}\left[Z^{(0)}\right]^2,
\end{align}
\end{subequations}
with $Z^{(0)}(-\infty)=1$ and $Z^{(1)}(-\infty)=0$. The zeroth-order solution is
\begin{equation}
    Z^{(0)}(\tau)=\exp\left[-\frac{v^2}{\alpha}\left(\arctan(\alpha\tau)+\frac{\pi}{2}\right)\right].
\end{equation}
Writing $r=v^2/\alpha$, we solve the first-order equation by an integrating-factor method. At the end of the sweep, this gives
\begin{equation}
    Z_{0}(+\infty)=e^{-\pi r}+\frac{2c}{\Gamma}\frac{r}{r^2+4}\left(e^{-\pi r}-e^{-2\pi r}\right)+\mathcal{O}\left(\frac{c^2}{\Gamma^2}\right).
\end{equation}
Using $\bar{P}=[1+Z_{0}(+\infty)]/2$ then recovers Eq.~(\ref{Eq-Phase-Strong}) in the main text.

\end{document}